\documentclass[journal]{IEEEtran}

\ifCLASSOPTIONcompsoc
\usepackage[nocompress]{cite}
\else
\usepackage{cite}
\fi

\ifCLASSINFOpdf
\else
\fi

\usepackage{url}
\usepackage{bm}
\usepackage{amsmath}
\usepackage{graphicx}
\usepackage{color}

\begin{document}

%
%
\catcode`\@=11
\newcommand{\gsim}{${\mathrel{\mathpalette\@versim>}}$}
\newcommand{\lsim}{${\mathrel{\mathpalette\@versim<}}$}
\newcommand{\@versim}[2]{\lower 2.9truept \vbox{\baselineskip 0pt \lineskip
    0.5truept \ialign{$\m@th#1\hfil##\hfil$\crcr#2\crcr\sim\crcr}}}
\catcode`\@=12
\def\be{\begin{equation}}
\def\bea{\begin{eqnarray}}
\def\ee{\end{equation}}
\def\eea{\end{eqnarray}}
\def\deg{$^\circ$}
\def\bs{\boldsymbol}
\def\arcmin{$^\prime$}
\def\arcsec{$^{\prime \prime}$}

\title{Model for a Noise Matched Phased Array Feed}

\author{D.~Anish~Roshi,~\IEEEmembership{Member,~IEEE,}
         W.~Shillue,~\IEEEmembership{Member,~IEEE,}
        and~J.~Richard~Fisher,~\IEEEmembership{Member,~IEEE}
\thanks{D. A. Roshi was with the National Radio Astronomy Observatory (NRAO), Charlottesville.
He is currently with the Arecibo Observatory, Arecibo, PR 00612. W.~Shillue and J. R. Fisher are with the NRAO, 520 Edgemont Road, Charlottesville, VA 22903 USA
e-mail: aroshi@naic.edu, bshillue@nrao.edu, rfisher@nrao.edu. }
\thanks{The National Radio Astronomy Observatory is a facility of
the National Science Foundation operated under a cooperative
agreement by Associated Universities, Inc. }
\thanks{Manuscript received ; revised .}}

\markboth{{IEEE} Transactions on Antennas \& Propagation,~Vol.~XX, No.~X, December~XXXX}%
{Roshi, Shillue \& Fisher : PAF model}

\maketitle

\begin{abstract}
We present a model for a Noise Matched Phased Array Feed (PAF) system 
and compare model predictions with the measurement results.
The PAF system consists of an array feed, a receiver, a beamformer 
and a parabolic reflector. The novel aspect of our model is
the characterization of the {\em PAF system} by a single matrix.
This characteristic matrix is constructed from 
the open-circuit voltage covariance at the output
of the PAF due to signal from the observing source,
ground spillover noise, sky background noise and
(low-noise) amplifier (LNA) noise. The best signal-to-noise ratio 
on the source achievable with the PAF system will be 
the maximum eigenvalue of the characteristic matrix.
The voltage covariance due to signal and spillover noise
are derived by applying the Lorentz reciprocity theorem. 
The receiver noise covariance and noise temperature are obtained 
in terms of Lange invariants such that they are  
suitable for noise matching the array feed.   
The model predictions
are compared with the measured performance of a 1.4 GHz, 19-element, dual-polarized PAF on the 
Robert C. Byrd Green Bank Telescope.\footnote{
The Green Bank Observatory is a facility of the National Science Foundation
operated under a cooperative agreement by Associated Universities, Inc.} 
We show that the model predictions, obtained 
with an additional noise contribution due to the measured losses ahead of the low-noise
amplifier, 
compare well with the measured ratio of system temperature to aperture efficiency  
as a function of frequency and as a function of offset from
the boresight. Further, our modeling indicates that the bandwidth over which this ratio is optimum
can be improved by a factor of at least two by noise matching the PAF with the LNA. 
\end{abstract}


\IEEEpeerreviewmaketitle

\section{Introduction}

\IEEEPARstart{T}{he} extreme faintness of celestial radio sources motivates the use of
telescopes with large collecting areas. The field-of-view (FoV) of a single-dish telescope 
that uses a single feed to receive signals decreases in proportion to the collecting area.
Traditionally, array feeds consisting of multiple horn antennas have been used to
increase the FoV of such telescopes. The antennas in such arrays cannot be closely packed
since the feeds are physically large as they have to illuminate the reflector
efficiently.  Consequently, the telescope beams corresponding to the different elements are separated by
considerably more than the half-power
beamwidth (HPBW). The non-overlapping beams reduce the mapping efficiency.
Feeds consisting of dense, electrically small antenna arrays
as a means to overcome this limitation are now of significant interest
\cite{fisherbradley2000, oosterlooetal2010, veidtetal2011, cortesmedellin2015, chippendaleetal2015, 
wuetal2016, haybird2015, warnicketal2016}. These dense arrays sample the focal field pattern
of the telescope. Multiple beams are then formed by combining the
signals sampled by the array elements with complex weights that form
an efficient reflector illumination.
Such arrays are referred to as Phased Array Feeds (PAF). The beams
formed using a PAF can fully sample the FoV. 
Additionally, a PAF can be used to improve spillover efficiency as well as the
illumination of the dish. However, mutual coupling between array elements is a major hurdle in designing
a low-noise PAF (see, for example, \cite{craeyeovejero2011}).  
Mutual coupling modifies the element radiation patterns.
It also couples amplifier noise between signal paths. 
Therefore, detailed electromagnetic, noise and network
modeling are needed to design a PAF for radio astronomy applications \cite{rfisher1996}.

Several research groups have analyzed and modeled the noise performance \cite{
warnickjensen2005, maaswoes2007, warnickjensen2007, hayosullivan2008, warnicketal2009, maas2010} and 
electromagnetic properties of a PAF on a reflector antenna
\cite{warnicketal2011, hay2010}.
A working expression for the covariance matrix of the signal due to a source 
at the output of the PAF was provided by Warnick \& Jeffs (2008) \cite{warnickjeffs2008} 
and corresponding matrix for
spillover noise was derived by Warnick \& Jeffs (2006) \cite{warnickjeffs2006}
(see also \cite{warnicketal2010}). These matrices were 
then used to provide expression for aperture and spillover 
efficiencies\cite{warnickjeffs2006,warnickjeffs2008}.  
The receiver noise covariance matrix for a PAF was derived by 
Warnick \& Jensen (2005) \cite{warnickjensen2005} (see also \cite{maaswoes2007}).  
Hay (2010) \cite{hay2010} (see also \cite{hayosullivan2008}) analyzed the noise performance of a
PAF with a lossless input matching network in terms of Lange invariants.
He also provided optimum solutions for the matching network for maximum signal-to-noise
ratio (SNR) and total efficiencies (see \cite{warnicketal2009} and references therein
for other optimization). Finally, simplification of the PAF receiver noise model
has been discussed by Ivashina et al. (2008) \cite{ivashinaetal2008} to better understand the 
factors affecting system sensitivity.

In this paper, we apply the Lorentz reciprocity theorem to a PAF mounted at the
prime focus of a reflector telescope to derive a new expression 
for the signal covariance matrix. 
We also provide a new expression for the 
receiver noise covariance matrix and noise temperature 
in terms of the Lange invariants, which are 
advantageous for numerically optimizing the system performance 
with a lossless matching network (see also \cite{belostotskietal2015}). 
A novel aspect of
our modeling is the introduction of a single matrix, referred to as the
characteristic matrix of the {\em PAF system}, constructed 
from these covariance matrices to express the SNR at the output of the PAF.
The PAF system 
is described in Section~\ref{pafsystem}.
A summary of the novel aspects of our PAF model and its implementation 
are presented in Section~\ref{pafmodel}. 
We then compare the model predictions with measurements
made with a 1.4 GHz, 19-element dual-polarized PAF on the 
Green Bank Telescope (GBT) \cite{roshietal2018}. The results
of the comparative study are given in Section~\ref{compmeas}. The main results of 
the paper are summarized in the Section~\ref{conclu}. A list of mathematical symbols used
in the paper is given in Appendix~\ref{symbol_list}. 

 

\section{The PAF system}
\label{pafsystem}

\begin{figure}[!t]
\centering
\begin{tabular}{c}
\includegraphics[width=1.5in]{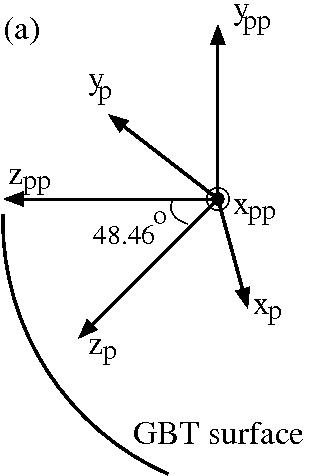} \\
\includegraphics[width=3.3in]{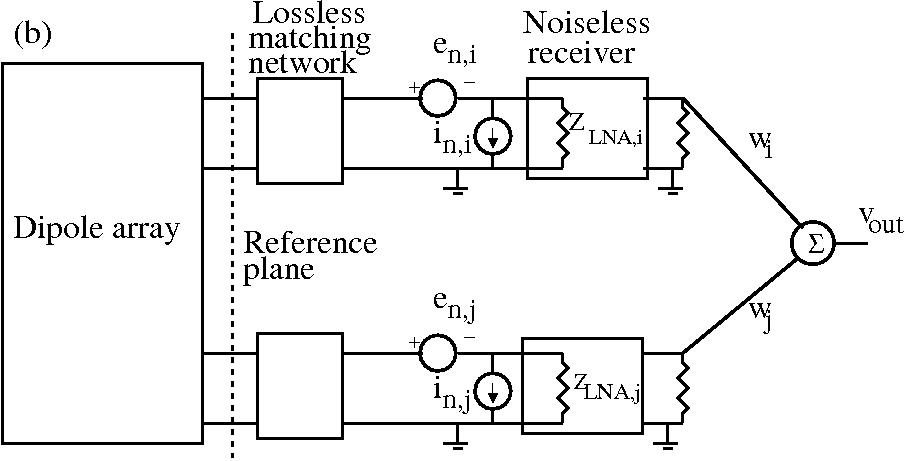}
\end{tabular}
\caption{The PAF system. (a) A schematic of the GBT reflector geometry.
The equation of the reflector surface is expressed in the $x_{pp}-y_{pp}-z_{pp}$ coordinate
system. { The $x_p-y_p-z_p$ coordinate system is used to describe the
properties of the PAF. The origins of the two coordinate systems coincide and
are located at the prime focus of the reflector.
The orientation of the $x_p-y_p-z_p$ coordinate system relative to the $x_{pp}-y_{pp}-z_{pp}$
is obtained through two rotations: a rotation about $x_p$ by 48.46\deg\ as shown and a rotation
about $z_p$ by 45\deg. The ground plane of the PAF coincides with the $x_p-y_p$ plane
and the X and Y polarization dipoles are parallel to $x_p$ and $y_p$ axes respectively. } 
(b) The PAF, noise matching network,
receiver system and the beamformer.  The noise voltage and current fluctuations
are related to the noise parameters $R_{n_i}$, $g_{n_i}$ and $\rho_i$ of the
LNA: 
$\langle e_{n,i}^2\rangle  =  4 k_B T_0 R_{n_i}$,  
$\langle i_{n,i}^2\rangle  =  4 k_B T_0 g_{n_i}$, and
$\rho_i   =  \frac{\langle e_{n,i}^*i_{n,i}\rangle}{\sqrt{\langle e_{n,i}^2\rangle \langle i_{n,i}^2\rangle}}$, where $k_B$ is the
Boltzmann constant and $T_0 = 290$ K. 
}
\label{fig1}
\end{figure}

\begin{figure}[!t]
\centering
\includegraphics[width=3.0in]{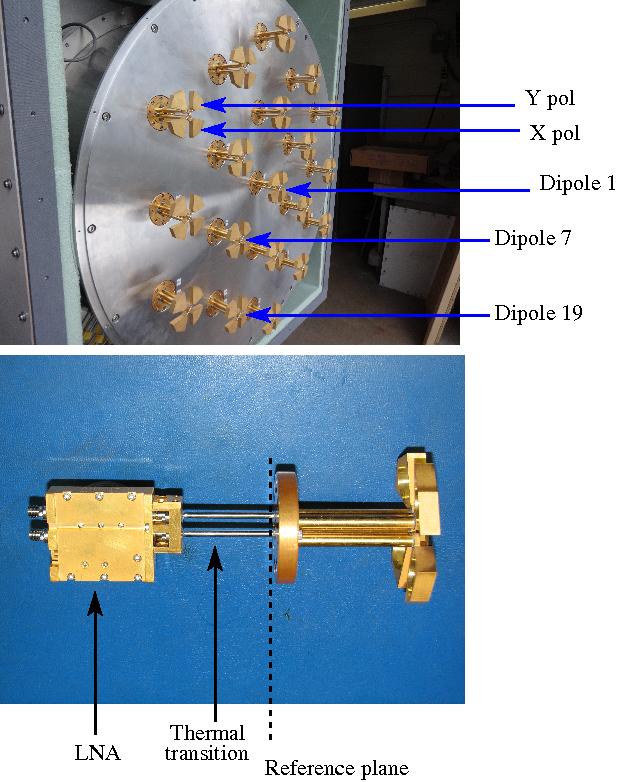}
\caption{The 19-element dual-polarized dipole array is shown on the top.
The array is designed to operate near 1350 MHz with a bandwidth $\ge$ 150 MHz.
The dipoles are numbered counter clockwise along each `ring'. The
dipoles designation for the X and Y polarizations are marked.
A dual polarized dipole pair
along with the thermal transitions and LNAs are shown in the bottom. The reference
plane marked on the figure corresponds to that shown in Fig.~\ref{fig1}. The
length of the thermal transition (6.4 cm) is fixed for the actual array, but
we have varied its length in the PAF modeling program to noise match the array to
the LNA.
}
\label{fig2}
\end{figure}

\begin{figure}[!t]
\centering
\begin{tabular}{c}
\includegraphics[width=3.0in]{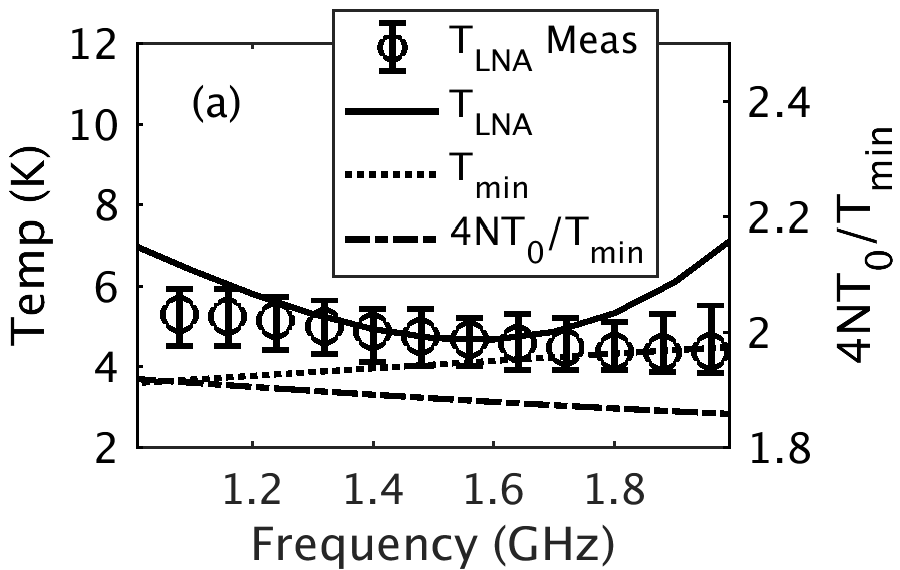} \\
\includegraphics[width=3.0in]{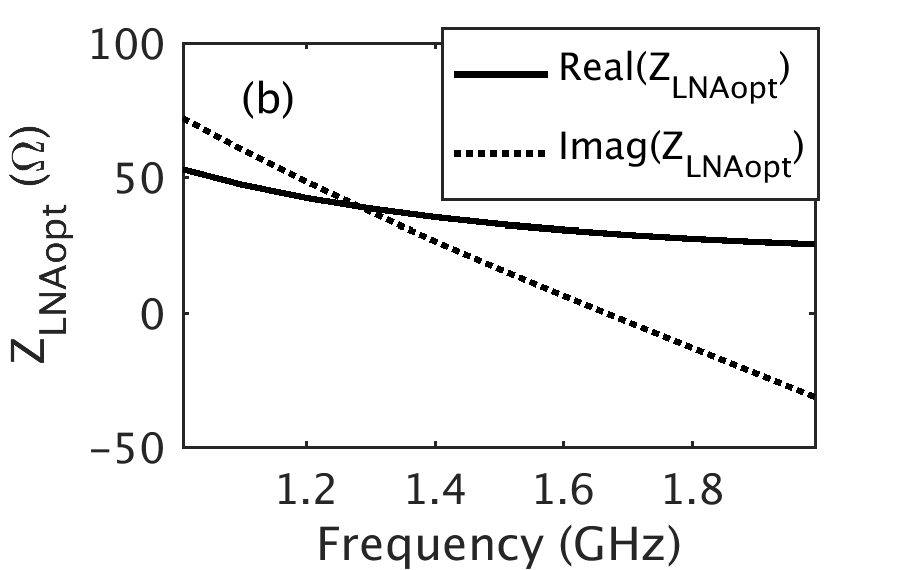}
\end{tabular}
\caption{(a)The measured mean noise temperature vs frequency of the LNA and its peak-to-peak scatter
are marked in circles with `error bar'. The mean (4.85 K) and scatter (1.4 K) are obtained 
from the measurements
of 38 LNAs. The noise temperature prediction from the noise model of the LNA is shown with
continuous line. The noise parameter $T_{min}$ and $4NT0/T_{min}$
are plotted in dotted and dashed lines respectively. (b) The real (continuous line)
and imaginary parts (dotted line)
of the $Z_{LNAopt}$ are plotted as a function of frequency.
}
\label{fig3}
\end{figure}

The PAF system consists of a dual-polarized dipole array followed by noise matching
networks, LNAs and receiver system, a beamformer and a parabolic reflector or telescope 
(see Fig.~\ref{fig1}). In Fig.~\ref{fig1}a,
we show an off-axis reflector, representing the surface of the GBT. 
The projected surface of the reflector in
the boresight direction (i.e perpendicular to the $x_{pp}-y_{pp}$ plane in Fig.~\ref{fig1}a) 
is circular with a diameter 100 m and the focal length over diameter ratio is 0.6. 
Further details of the GBT geometry are
available in \cite{hallking1992, norridsri1996}. 
The PAF is mounted at the prime focus with the
plane of the array at an angle 48.46\deg\ from the $x_{pp}-y_{pp}$ plane 
{ (see Fig.~\ref{fig1}a).}

The dipole array consists of 19 dual-polarized elements (see Fig~\ref{fig2}).  
The dipole shape was optimized
for active impedance match to the LNA and for
maximum sensitivity on the GBT over a FoV of $\sim$ 20 arcminutes \cite{warnicketal2011,
warnicketal2009}. The frequency range used for optimization was $\ge$ 150 MHz centered
at 1350 MHz. 
A balun converts the signals received
by the dipoles to unbalanced signals, which are available at the output of a transmission
line with characteristic impedance $z_0$, which is same as the 50 $\Omega$ reference impedance. 
The transmission line terminates at the `reference plane' marked on Fig~\ref{fig1}b 
(see also Fig.~\ref{fig2}).
The dipole array, balun, ground plane and the transmission line together is 
referred to as the PAF. Further details of the PAF is available in \cite{roshietal2018}.

The PAF is followed by a two-port, lossless matching network, which in the simplest
case can be a transmission line of length $L_{trans}$.
The transmission line in the measurement system is the thermal transition (see Fig.~\ref{fig2})
with $L_{trans} \sim$ 6.4 cm and characteristic impedance of 50 $\Omega$.
The thermal transition is made of air-core stainless steel
co-axial transmission line with the central stainless steel conductor
coated with copper and gold of 5$\mu m$ thickness.
As discussed in Section~\ref{compmeas}, we have used this transmission
line as the matching network in the PAF model and have varied $L_{trans}$
for all dipoles between 3 and 20 cm for noise matching the PAF to the LNA.
The LNAs are located at the end of the matching network. 
The LNAs used for the measurements were cryogenic Silicon Germanium Heterojunction 
Bipolar Transistor (SiGe HBT) amplifiers cooled to about 15 K \cite{grovesmorgan2017, roshietal2018}. 
The $i^{th}$ LNA has an input impedance of $Z_{LNA_i} \sim 50\; \Omega$ and
gain $\sim 38$ dB. The input impedance of the receiver system is transformed by 
the matching network to $Z_{in_i}$ at the reference plane. 
The noise properties of the LNA are usually specified by
the  noise parameters $R_{n_i}$, the noise resistance,  $g_{n_i}$, the noise conductance,
and $\rho_i$, the complex noise correlation \cite{pospieszalski2010} (see also Fig.~\ref{fig1}b).
An equivalent set of noise parameters used
in radio astronomy are $Z_{LNAopt_i}$, the optimum impedance and
the Lange invariants, $T_{min_i}$ (the minimum noise temperature) 
and $N_i$ \cite{lange1967, pospieszalski2010}. The relationship
between the two sets of noise parameters is discussed in \cite{pospieszalski2010}. The latter set of
noise parameters is advantageous when lossless networks are used for noise matching. This is
because the noise parameters $T_{min_i}$ and $N_i$ remain invariant when
they are transformed through a lossless network placed at the input of the LNA and $Z_{LNAopt_i}$
transforms to $Z_{opt_i}$ like a normal impedance transformation.  
Fig.~\ref{fig3} shows the noise parameters vs frequency obtained from the noise model of the LNA.
The predicted temperature from the noise model of the LNA is consistent with measurements
in the frequency range 1.2 to 1.7 GHz (see Fig.~\ref{fig3})\cite{grovesmorgan2017}. 
However, the model noise
temperature deviates from measurement outside this frequency range. We are in the process
of improving the noise model of the LNA and the results will be presented elsewhere.

In the measurement system, the signals from the output of the LNAs 
are further amplified, down-converted to basebands, digitized using 
analog-to-digital converters (ADC) and then followed by   
Fast Fourier Transform (FFT) engines to create time series of complex voltage spectra
(see \cite{roshietal2018} for a full description of the system). 
The basic signal processing done in a PAF system, referred to as {\em beamforming},
can be expressed as \cite{jeffsetal2008}
\be
v_{out}  =  \sum_{i=1,M} w_i v_i  =  \bm w^T \bs{\tilde{V}}
\ee
where the elements of the voltage vector $\bs{\tilde{V}}$ are the 
input complex voltages $v_i$ from a spectral channel to the beamformer, the elements of the 
weight vector $\bm w^T$ are the complex beamforming weights $w_i$ and
$M = 19 \times 2$ is equal to the number of elements in the PAF. 
Multiple beams are formed by applying different sets of beamforming weights. 

For modeling the PAF, its network property is specified by the (spectral) impedance matrix $\bm Z$
and its electromagnetic property is specified by the {\em embedded beam patterns} \cite{rudgeetal1983} 
(see also \cite{roshifisher2016, roshi2017}). 
In this paper, we define the $i^{th}$ embedded beam pattern,
$\vec{\mathcal{\psi}}^e_i(\vec{r})$, as the radiation pattern of the PAF when $i^{th}$ port 
is excited with 1 A and all other ports are open circuited. 
The position vector $\vec{r}$ is defined in the $x_p-y_p-z_p$ coordinate system (see
Fig.~\ref{fig1}a). There will be $M = 19 \times 2$ embedded beam patterns for the dual
polarized array. 
The beam pattern at the far-field can be expressed as an out-going spherical wave; i.e. 
$\vec{\mathcal{\psi}}^e_i(\vec{r}) =  \vec{\Psi}^e_i(\theta, \phi) \; \frac{e^{j\vec{k}.\vec{r}}}{r}$,
where $\vec{k}$ is the propagation vector and $r = |\vec{r}|$.
The radiation pattern of the PAF when excited by a set of arbitrary
port currents $\mathcal{J}_{0_i}$ is 
\be
\vec{\mathcal{E}}(\vec{r})  =   \sum_{i=1,M} \mathcal{J}_{0_i} \vec{\mathcal{\psi}}^e_i(\vec{r}) 
                            =   \bm I_0^T \bs{\vec{\mathcal{\psi}}^e},
\label{PAFfpat2}
\ee
where $\bm I_0$ is the vector of port currents and $\bs{\vec{\mathcal{\psi}}^e}$ is
the vector of embedded beam patterns. 
The $\theta, \phi$ dependence of the far-field beam pattern can be written in
a similar fashion: 
$\vec{E}(\theta, \phi) = \bm I_0^T \bs{\vec{\Psi}^e},$
where $\bs{\vec{\Psi}^e}$ is a vector with elements $\vec{\Psi}^e_i$.
The unit of $\vec{\mathcal{E}}$ is V/m, that of $\vec{\mathcal{\psi}}^e_i$ is V/A/m, 
that of $\vec{E}$ is V and that of $\vec{\Psi}^e_i$ is V/A.

The voltages, currents and field amplitudes are considered harmonic quantities 
and their values are specified as {\em peak} values ($e^{j\omega t}$ term is 
omitted for simplicity) for the derivation of the open circuit voltage 
(Eq.~\ref{voc}) by applying Lorentz reciprocity theorem. When applying
this result to the PAF system, we need to consider radiation field from
the astronomical and thermal sources. These fields and the induced
voltages at the output of the PAF as well as the noise due to the LNA 
are stochastic quantities. We treat these signals in the quasi monochromatic 
approximation and their amplitudes are taken as RMS (root mean square) 
values. 
{
In this treatement, for example, the flux density of a stochastic radiation field $\vec{E}_{inc}$
is $S = \frac{1}{Z_0}\langle \vec{E}_{inc}\cdot\vec{E}_{inc}^* \rangle$ and
will have units W/m$^2$/Hz. 
}
The open circuit voltage covariance 
$\langle \bm V_{oc} \bm V_{oc}^H \rangle$, where
$\bm V_{oc}$ is the induced voltage due to stochastic field or noise,
is average power dissipated by the noise in unit conductance 
and unit bandwidth or W/Hz/$\Omega$. 

\section{PAF model}
\label{pafmodel}

One of the novel aspects of our PAF model is the introduction of a new matrix,
referred to as the characteristic matrix, to concisely represent the
PAF system. Further, we apply the Lorentz reciprocity 
theorem \cite{clarkebrown1980} to derive a new expression for the signal 
covariance matrix. The receiver covariance matrix is obtained 
in terms of the Lange invariants so that it is suitable for
noise matching the array feed to the LNA. 
We summarize below these new aspects of the PAF theory and
also briefly describe its implementation. 

\subsection{PAF Theory Revisited}
\label{theory}

The question posed to develop our model for a lossless PAF is:
what is the maximum signal-to-noise ratio that can be obtained
with a PAF system when observing a compact source (point source)
at some angle $\theta_s$, $\phi_s$ from the boresight direction ?
The answer to this question is given by the following theorem:
\subsubsection*{Theorem} Given (a) the (spectral) impedance matrix, $\bm Z$,
and the embedded beam patterns, $\bs{\vec{\mathcal{\psi}}^e}$, of the PAF,
(b) the LNA noise parameters, $T_{min_i}$, $N_i$ and the transformed
impedance, $Z_{opt_i}$ and $Z_{in_i}$, (c)
the telescope geometry and ground temperature, 
(d) the flux density and direction $\theta_s$, $\phi_s$ of the observing
source and (e) the off-source sky temperature, one can
construct a characteristics matrix $\bm M$ for the {\em PAF system}.
Then the best signal-to-noise ratio on the source is the maximum eigenvalue, 
$e_{max}$, of $\bm M$.

\subsubsection{Characteristic Matrix and Proof of the Theorem}

The SNR when observing a source with the PAF system is defined 
as the ratio of increase in power spectral
density at the output of a beam due to the source relative to the
off-source spectral density (see Eq. 6 \cite{warnickjeffs2008, jeffsetal2008}). 
This SNR at the beamformer output can be expressed in terms
of the covariance of the open circuit voltages at the 
reference plane shown in Fig.~\ref{fig1} as 
\be
\textrm{SNR}   =  \frac{\bm w_1^H \bm R_{signal} \bm w_1}{\bm w_1^H \bm N \bm w_1} 
               =  \frac{\bm w_2^H \bm M \bm w_2}{\bm w_2^H \bm w_2}
\label{snr1}
\ee
where $\bm R_{signal}$ is the open circuit covariance matrix due to signal from the source, 
\be
\bm N  = \bs{\mathcal{N}} \bs{\mathcal{N}} \equiv \bm R_{spill} + \bm R_{rec} + \bm R_{sky}
\label{nmatrix}
\ee
is the sum of the open circuit noise covariance matrices due to spillover,$\bm R_{spill}$, receiver
$\bm R_{rec}$, and sky background radiation, $\bm R_{sky}$, 
\be
\bm M \equiv \bs{\mathcal{N}^{-1}} \bm R_{signal}  \bs{\mathcal{N}^{-1}},
\label{charmat}
\ee
is defined as the characteristics matrix. The relationship between the weight vectors
are $\bm w_1  =  
(\bm  Z_{in}(\bm  Z + \bm  Z_{in})^{-1})^H \bm G^H \bm w$
and $\bm w_2 = \bs{\mathcal{N}} \bm w_1$,  
$\bm  Z_{in}$ is a diagonal
matrix of elements $Z_{in_i}$  and $\bm G$ is a diagonal matrix with
elements as the overall system gain. 
It follows from Eq.~\ref{snr1} that the maximum SNR is given by the 
maximum eigenvalue, $e_{max}$, of $\bm M$. The weight vector that needs
to be applied in the beamformer can be obtained from the eigenvector
corresponding to $e_{max}$. 

\subsubsection{Open circuit voltage covariance}

Application of the Lorentz reciprocity theorem provides an expression for 
the open circuit voltage $\bm V_{oc}$ at the output
of the PAF as \cite{roshifisher2016, roshi2017} 
\be
\bm V_{oc} = \int_{A_{free}} \left(\bs{\vec{\mathcal{\psi}}^e}^T
             \times \bs{\mathcal{I}} \vec{\mathcal{H}_r} -
             \bs{\mathcal{I}} \vec{\mathcal{E}_r} \times \bs{\vec{\mathcal{J}}^e}\right)
             \cdot \hat{n}\; \textrm{d}A. 
\label{voc}
\ee
Here $\bs{\vec{\mathcal{J}}^e}$ is the vector of magnetic
field patterns corresponding to the embedded beam pattern $\bs{\vec{\mathcal{\psi}}^e}$,
$\vec{\mathcal{E}_r}$ and $\vec{\mathcal{H}_r}$ are the incident electric and magnetic 
fields respectively, $\bs{\mathcal{I}}$ is the identity matrix and $\hat{n}$ is the 
unit normal inward to the elementary area $\textrm{d}A$. 
The surface of integration $A_{free}$ is part of a closed surface outside the PAF.

\subsubsection{$\bm R_{signal}$}

We consider a compact astronomical source in the direction $\theta_s$,  $\phi_s$
from the boresight.
For simplicity we assume that the source is unpolarized, which implies the
following relationship between the flux density, $S_{source}$,  and electric field 
of the incident plane wave, $\vec{E}_{inc}$, due to the source:
$\frac{S_{source}}{2} = 
\frac{1}{Z_0}\langle E_{inc,x}E_{inc,x}^* \rangle = 
\frac{1}{Z_0}\langle E_{inc,y}E_{inc,y}^* \rangle $, where $E_{inc,x}$
and $E_{inc,y}$ are the two linearly polarized components of $\vec{E}_{inc}$
and $Z_0$ is the free space impedance. An appropriate closed surface, with the
projected aperture plane (i.e. the telescope aperture plane projected onto a plane
perpendicular to the direction of the source) as part of this surface can be considered for the
evaluation of the integral in Eq.~\ref{voc}. The embedded beam patterns are
propagated to the projected aperture plane for the evaluation of the integral
(see Section~\ref{embimpm}). 
The major contribution to the integral over $A_{free}$ comes from the projected aperture 
plane \cite{roshifisher2016}. The open circuit voltage covariance 
due the source can be obtained as  
\be
\bm R_{signal}  = \frac{2 S_{source}}{Z_0} \bm C_{I\psi} 
\label{signal1}
\ee
where 
\be
\bm C_{I\psi} \equiv \left(\int_{A_{pap}} \bs{\vec{\mathcal{\psi}}^e_{pap}} \textrm{d} A \right) \cdot
                  \left(\int_{A_{pap}} \bs{\vec{\mathcal{\psi}}^e_{pap}} \textrm{d} A \right)^H.
\label{forsig}
\ee
Here $\bs{\vec{\mathcal{\psi}}^e_{pap}}$ is the vector of the
propagated embedded beam pattern on the projected aperture plane $A_{pap}$.
Note that $\bm C_{I\psi}$ is a function of source position $\theta_s, \phi_s$.
The unit of $\bm R_{signal}$ is power spectral density per
unit conductance.

We also provide an expression for the antenna temperature, $T_A$, which is 
a generalization of the equation for $T_A$ for a reflector antenna with a single feed. 
The power spectral density due to the source, which is proportional to 
$\bm w_1^H \bm R_{signal} \bm w_1$, 
can be expressed as a physical temperature by comparing it with
the power spectral density when the PAF system is at thermal
equilibrium with the reference temperature $T_0$ 
\cite{kerrranda2010} (see also \cite{warnicketal2010}): 
\bea
T_A &  = & \frac{S_{source} A_{ap}}{2 k_B} \left( \frac{\bm w_1^H \bm C_{I\psi} \bm w_1}
           {A_{ap}\; \bm w_1^H \bm C_{C\psi}  \bm w_1} \right), \nonumber \\
    &  = & \frac{S_{source} A_{ap} \eta_{ap}}{2 k_B},
\label{tant}
\eea
where
\be
\bm C_{C\psi}  \equiv  \int_{4\pi} \bs{\vec{\Psi}^e}\cdot \bs{\vec{\Psi}^e}^H \textrm{d}\Omega,
\label{fortemp}
\ee
and the aperture efficiency, $\eta_{ap}$, is defined in terms of the overlap integrals of the
embedded beam patterns and the physical area of the telescope aperture, $A_{ap}$, as 
(see also \cite{warnickjeffs2008})
\be
\eta_{ap} \equiv \frac{\bm w_1^H \bm C_{I\psi} \bm w_1}
           {A_{ap}\; \bm w_1^H \bm C_{C\psi}  \bm w_1}.
\ee

\subsubsection{$\bm R_{spill}$}

An expression for the covariance matrix of the voltage due to
ground spillover radiation seen by the PAF is obtained 
in \cite[their Eq. 9]{warnickjeffs2008}.
Applying Eq.~\ref{voc} and following the arguments in \cite{warnickjeffs2008}
$\bm R_{spill}$ can be readily expressed in terms of 
ground temperature, $T_g$ (= 300 K), and the overlap integral of the embedded
beam patterns as \cite{roshifisher2016, roshi2017}
\be
\bm R_{spill}  = \frac{4 k_B T_g}{Z_0} \bm C_{c\psi1},
\label{spill5}
\ee
where
\be
\bm C_{c\psi1} \equiv \int_{\Omega_{spill}} \bs{\vec{\Psi}^e}\cdot \bs{\vec{\Psi}^e}^H \textrm{d}\Omega,
\label{spill6}
\ee
$\Omega_{spill}$ is the solid angle over which the PAF embedded beam patterns
are receiving radiation from ground. 
Eq.~\ref{spill5} differs by a factor of 2 compared to Eq. 9 in \cite{warnickjeffs2008}
because $\bm R_{spill}$ is the power spectral density per unit conductance
(see Section~\ref{pafsystem}) and 
\cite{warnickjeffs2008} provides the peak voltage covariance. 

The spillover temperature, $T_{spill}$, as in the case of antenna temperature 
discussed above, is a generalization of the equation for $T_{spill}$ for 
a reflector antenna with a single feed \cite{roshifisher2016, roshi2017}:
\be
T_{spill} =  T_g (1-\eta_{spill})
\ee
where 
\be
\eta_{spill} \equiv 1 - \frac{\bm w_1^H  \bm C_{C\psi1}  \bm w_1}{\bm w_1^H  \bm C_{C\psi}   \bm w_1}.
\label{etaspill}
\ee
In \cite{warnickjeffs2008}, Twiss's result \cite{twiss1955} 
has been re-written in terms of the peak voltage covariance and  
so Eq.~\ref{etaspill} and Eq. 16 in \cite{warnickjeffs2008} give
the same result.

\subsubsection{$\bm R_{rec}$}

The open circuit voltage covariance due to the LNA noise {\em in the absence of
a matching network} is given by \cite{warnicketal2009} 
\bea
\bm R_{rec} & =  & \langle \bm V_{oc_r} \bm V_{oc_r}^H \rangle \nonumber   \\
      & =  & \langle \bm E_n\bm E_n^H\rangle + \bm Z\; \langle \bm I_n \bm E_n^H \rangle + \nonumber   \\
      &    & \langle \bm E_n \bm I_n^H\rangle \; \bm Z^H +
             \bm Z \; \langle \bm I_n \bm I_n^H\rangle \; \bm Z^H.
\label{ww1}
\eea
Here $\langle \bm E_n \bm E_n^H\rangle$, $\langle \bm I_n \bm I_n^H\rangle$ and
$\langle \bm I_n \bm E_n^H\rangle$
are diagonal matrices of noise fluctuations $\langle e_{n,i}^2\rangle$, $\langle i_{n,i}^2\rangle$ 
and their correlations $\rho_i$ (see Fig.~\ref{fig1}b).
For identical LNAs connected to the PAF, i.e. $R_{n_i} = R_{n}$,
$g_{n_i} = g_{n}$, $\rho_i = \rho$, $T_{min_i} = T_{min}$,
$N_i = N$ and $Z_{LNAopt_i} = Z_{LNAopt}$ for all $i$, we get
\begin{multline}
\bm R_{rec} = 4 k_B T_0 \; \Big(R_n \bs{\mathcal{I}} + \\
 \sqrt{R_n g_n} \;\big(\rho \bm Z + \rho^* \bm Z^H\big) 
            + g_n \bm Z\bm Z^H \Big).
\label{tn}
\end{multline}
We re-write $\bm R_{rec}$ in terms of the Lange invariants,
and the optimum impedance (see Appendix~\ref{rec_temp_ap}),
\begin{multline}
\bm R_{rec}  =  4 k_B T_{min}\; \frac{\big(\bm Z + \bm Z^H\big)}{2} \\
         + 4 k_B T_0 N \;\frac{\big(\bm Z - Z_{LNAopt}\bs{\mathcal{I}}\big) \; 
             \big(\bm Z - Z_{LNAopt}\bs{\mathcal{I}}\big)^H}{{\rm Re}\{Z_{LNAopt}\}}.
\label{tn1}
\end{multline}
The unit of $\bm R_{rec}$ is power spectral density per
unit conductance.
When a matching network is introduced, $\bm R_{rec}$ at the reference plane
can be obtained by replacing $Z_{LNAopt}$ with the transformed impedance $Z_{opt}$
in Eq.~\ref{tn1}.  

The receiver temperature, $T_{rec}$, at the reference plane can be obtained 
from the power spectral density, which is proportional to $\bm w_1^H \bm R_{rec} \bm w_1$,
as (see also \cite{warnicketal2010}) 
\be
T_{rec} = T_{min} + N T_0 \; \frac{\bm w_1^H (\bm Z - Z_{opt} \bs{\mathcal{I}})
   (\bm Z - Z_{opt} \bs{\mathcal{I}})^H \bm w_1} {\mbox{Re}\{Z_{opt}\} \; \frac{1}{2}\bm w_1^H (\bm Z + \bm Z^H) \bm w_1}.
\label{youreq}
\ee
Eq.~\ref{youreq} is a generalization of the expression for $T_{rec}$ versus source impedance 
for a single antenna connected to a receiver (see Eq.~11 in \cite{pospieszalski2010}).

\subsubsection{$\bm R_{sky}$ and Losses in the PAF system}

The sky background radiation has components due to cosmic microwave background (CMB), galactic
and extragalactic radiation. We also include atmospheric radiation along with the sky contribution.
When comparing the
model results with measurement, it is also required to account for the losses
in the PAF system ahead of the LNA. These losses are not included in the PAF theory
and characterizing them is an ongoing research. In our model,
the open circuit voltage covariance matrices due to these components are taken
as \cite{twiss1955}
\be
\bm R \approx 2 k_B T (\bm Z + \bm Z^H).
\label{sky}
\ee
For $\bm R = \bm R_{sky}$, the covariance matrix due to sky radiation, 
$T= T_{cmb} + T_{atmo} + T_{bg,\nu_0} \left(\frac{\nu}{\nu_0}\right)^{-2.7}$,
where $T_{cmb} = 2.7$ K is the cosmic microwave background temperature, 
$T_{atmo} = 2$ K \cite{jean-yvesetal2002} is the atmospheric radiation temperature
(assumed to be frequency independent near 1.4 GHz) and   
$T_{bg,\nu_0} = 0.7$ K is the galactic background radiation
temperature at $\nu_0 = 1.42$ GHz \cite{reichreich1986} and $\nu$ 
is the frequency at which $\bm R_{sky}$
is computed. For $\bm R = \bm R_{loss}$, the covariance matrix due to losses,
$T=T_{loss}$, which is the physical temperature corresponding to the
noise due to losses in the
PAF system ahead of the LNA (see Section~\ref{loss}). When comparing 
the model results with the measurements we have added 
$\bm R_{loss}$ to $\bm N$ in Eq.~\ref{nmatrix}.




\subsection{The embedded beam pattern and the Impedance matrix}
\label{embimpm}

For the implementation of the theory presented above, 
the electromagnetic and network properties of the PAF were 
obtained using CST microwave studio (CST MWS\footnote{ 
https://www.cst.com/products/cstmws}). A 3D model of a
dual polarized element and balun were created in CST MWS
by importing the Autodesk Inventor\footnote{ 
https://en.wikipedia.org/wiki/Autodesk\_Inventor}
CAD model of the dipole assembly and the array was created by placing
copies of them. A rectangular ground plane of dimension
114.3 cm $\times$ 95.3 cm was added at a distance of 5.8 cm  
below the dipole elements. The full-wave 3D electromagnetic solver (transient
solver in CST MWS) was used to get radiation
patterns and scattering matrices of the PAF with 
hexahedral mesh size of $\lambda/20$.
The solver would excite ports one at a time
while all the other ports were terminated with the port impedance of 50 $\Omega$
to obtain the radiation pattern and scattering parameter. The final outputs 
of the CST simulation were these radiation patterns and scattering matrices for 
the different frequencies. 
The coordinate system that was used to obtain the radiation patterns
corresponded to the $x_p-y_p-z_p$ system defined in Section~\ref{pafsystem} (see Fig.~\ref{fig1}a). 
A MATLAB program was developed to compute the embedded beam patterns, as defined in
Section~\ref{pafsystem}, from the CST outputs. The impedance
matrix from the scattering parameter was also obtained using this program. 

The embedded beam patterns need to be propagated
to the projected aperture plane to compute $\bm R_{signal}$ (see Eq.~\ref{signal1}).
We used geometric optics approximation to obtain the fields in the aperture plane 
 \cite{yaghjian1984}. 
The reflected field due to each incident embedded 
beam pattern on the telescope surface is first computed after accounting
for the spherical spreading loss and then propagated to the aperture plane.
Note that the geometric phases due to the location of feed elements (or in other words 
the excitation current distribution) relative to the coordinate origin
were included in the computed embedded beam patterns (see \cite{roshifisher2016} for
further details).

\subsection{The losses ahead of the LNA}
\label{loss}

The receiver temperature of the PAF was measured in the outdoor test facility at
the Green Bank Observatory (GBO) \cite{roshietal2018}. The hot/cold load method 
was used to measure the receiver temperature \cite{warnicketal2010, 
woestenburg2012, chippendaleetal2014}. A
warm absorber, placed close to the dipole array, formed the hot load. The array
was pointed vertically to observe a cold sky region, which formed the cold load.
The effect of ground scattered radiation was mitigated to some extent by placing a 
metallic cone around the array during measurement. 
These measurements provide the receiver noise covariance matrices for
different frequencies which are then converted to  
temperature covariance matrices using the hot and cold load measurements \cite{roshietal2018}. 
The off-diagonal terms in the receiver temperature matrix have contributions from the 
mutual coupling in the array and the residual ground scattered radiation. 
These correlations can be canceled out to a large extent by taking the minimum eigenvalue 
of the temperature covariance matrix. This minimum eigenvalue we 
refer to as the minimum receiver temperature, $T_{rec,min}$.
The measured $T_{rec,min}$ as a function of frequency is shown in
Fig.~\ref{fig9}. The minimum difference of $\sim$ 2.5 K between $T_{rec,min}$ and 
the measured amplifier noise temperature provides an upper limit to the losses ahead 
of the LNA \cite{roshietal2018}. The contributions to the uncertainty in the measured upper limit
include (a) thermal uncertainty of the measurement ($\sim$ 0.3 K), 
(b) uncertainty in the LNA noise measurement ($\sim$ 0.3 K) and 
(c) uncertainty in the measured receiver temperature due to error in the hot and
cold load temperatures ($\sim$ 1 K). Thus the net 1$\sigma$ uncertainty 
of this upper limit is $\sim$ 1.1 K. 

\subsection{A computational model}
\label{modoutput}

A MATLAB\footnote{https://www.mathworks.com/} 
program was developed to apply the theory presented in
Section~\ref{theory} to the PAF system described in Section~\ref{pafsystem}.
The program starts with the computation of
$\bm R_{rec}$, $\bm R_{sky}$, $\bm R_{spill}$ and $\bm R_{signal}$. 
These matrices are computed for a set of frequencies where impedance matrices and
embedded beam patterns are available. $\bm R_{signal}$ is computed for a set of 
directions to the source within the desired FoV. The characteristic matrix $\bm M$ is 
then computed after adding $\bm R_{loss}$ to the noise matrix. 
The maximum SNR and the corresponding
weight vectors are obtained from $\bm M$ for the set of frequencies and positions. 
The measured performance of the PAF system
is usually expressed in $T_{sys}/\eta$, where $T_{sys}$ is the system temperature
and $\eta$ is approximately the aperture efficiency $\eta_{ap}$ of the telescope
(assuming that the radiation efficiency of the PAF is close to unity). 
The model maximum SNR is converted
to $T_{sys}/\eta$ using the flux density of the source and the physical area of the
telescope aperture \cite{roshietal2018}. 
The flux density models (typical uncertainty in flux density $<$ 5\%) 
of the observed sources are taken from Perley and Butler (2017)
 \cite{perleybutler2017}.  

The model also provides the full polarization field patterns in the 
aperture plane and the far-field pattern of the telescope. These are provided for
a specified set of weights. Further, the model computes the antenna temperature,
aperture efficiency, spillover temperature and spillover efficiency for
the specified weights.

\section{Comparison with measurement}
\label{compmeas}

The performance of a cryogenic 19-element dual-polarized PAF was 
measured on the GBT in March 2017 \cite{roshietal2018}. 
The observations that were made can be categorized into two groups. (1) Measurement of
the performance of the boresight beam as a function of frequency. For this
measurement, on-source and off-source voltage covariance on a set of calibrators
were obtained. The covariances were measured for a set of frequencies in 
the range 1200 to 1500 MHz, each averaged over a 
bandwidth of $\sim$ 300 KHz.  
(2) Measurement of the system performance over the FoV. Voltage covariances 
on a grid of positions centered on a strong calibrator source Virgo A were 
obtained for these measurements at 1336 MHz. 

The SNR on a source was obtained from the on-source 
and off-source measurements for both classes of observations. 
The beamformer weights were obtained by 
maximizing the SNR. The maximum SNR is expressed 
as the ratio of system temperature over efficiency, $T_{sys}/\eta$. 
The current measurements, however, do not provide separate values for $T_{sys}$ and
aperture efficiency. 

The Y-polarization data set were affected by two faulty signal paths and
telescope pointing offset. For comparison here, only a subset
of Y-polarization data that are not severely affected by these problems are used
(see \cite{roshietal2018} for further details). 

\subsection{Boresight $\frac{T_{sys}}{\eta}$}

\begin{figure}[!t]
\centering
\includegraphics[width=3.0in]{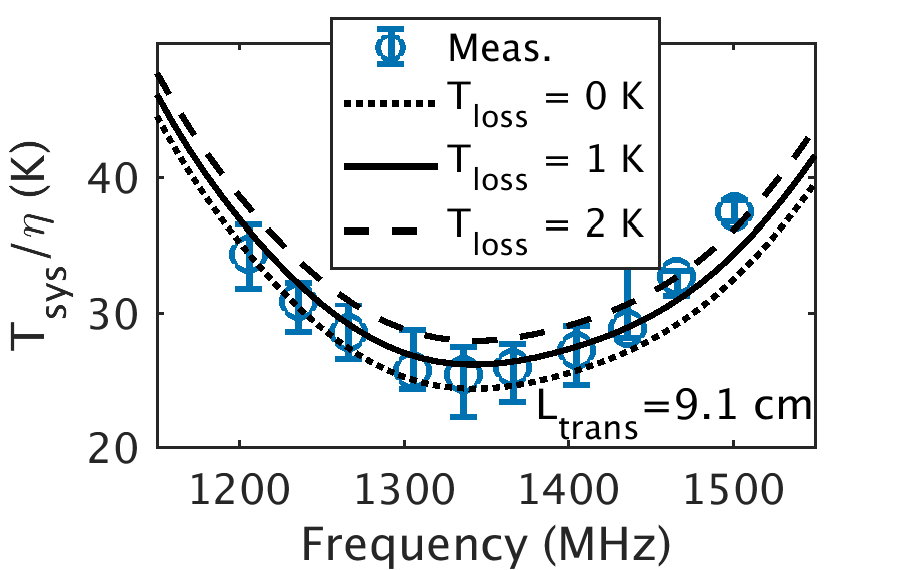} 
\caption{The modeled boresight $T_{sys}/\eta$ vs frequency for the
X polarization compared with the measured values taken from Roshi et al. (2018) \cite{roshietal2018}. 
The median of measured values with peak-to-peak variation are shown
in circles with `error bar'. The PAF model results for thermal transition
length (used as matching network in the model) $L_{trans}$ = 9.1 cm are plotted for 
$T_{loss} = 0, 1$ and 2 K. 
}
\label{fig5}
\end{figure}

\begin{figure}[!t]
\centering
\begin{tabular}{c}
\includegraphics[width=3.0in]{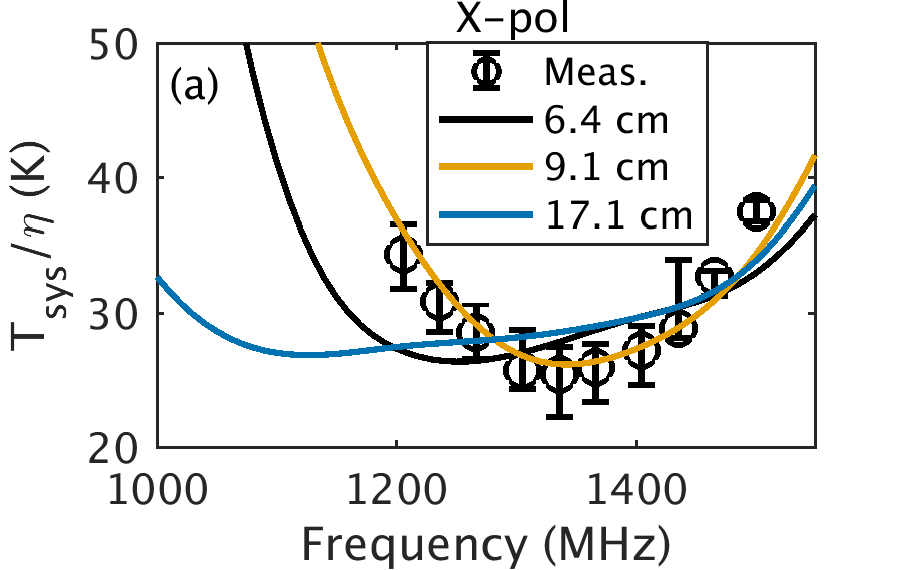} \\
\includegraphics[width=3.0in]{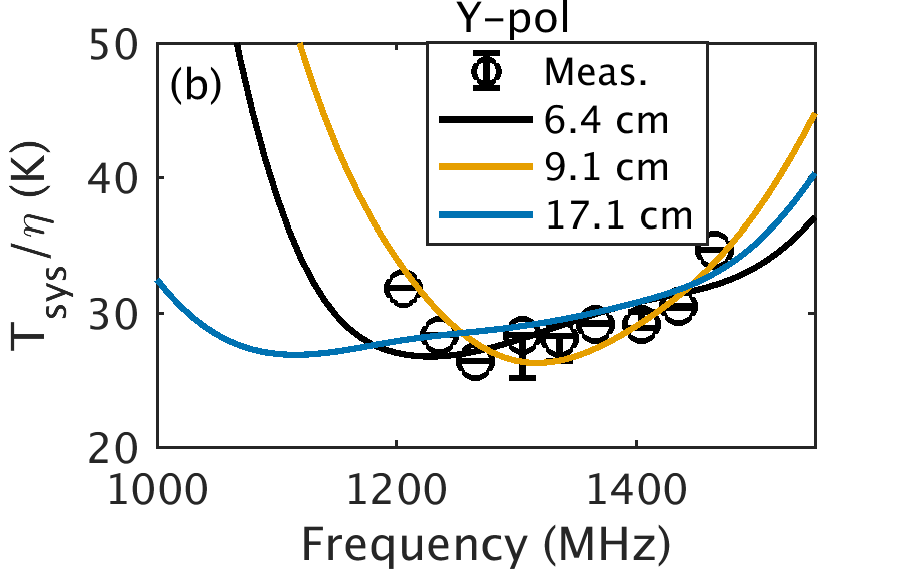} \\
\includegraphics[width=3.0in]{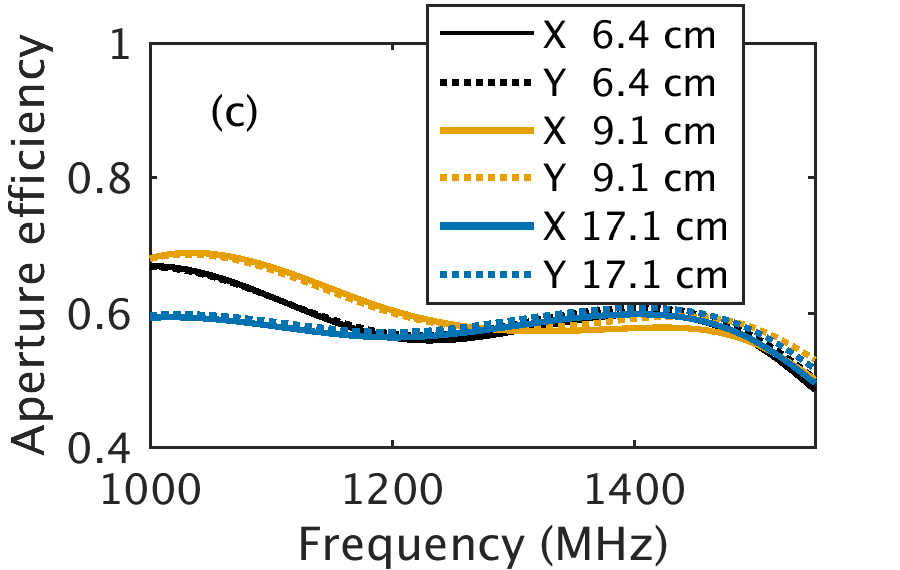} \\
 \includegraphics[width=3.0in]{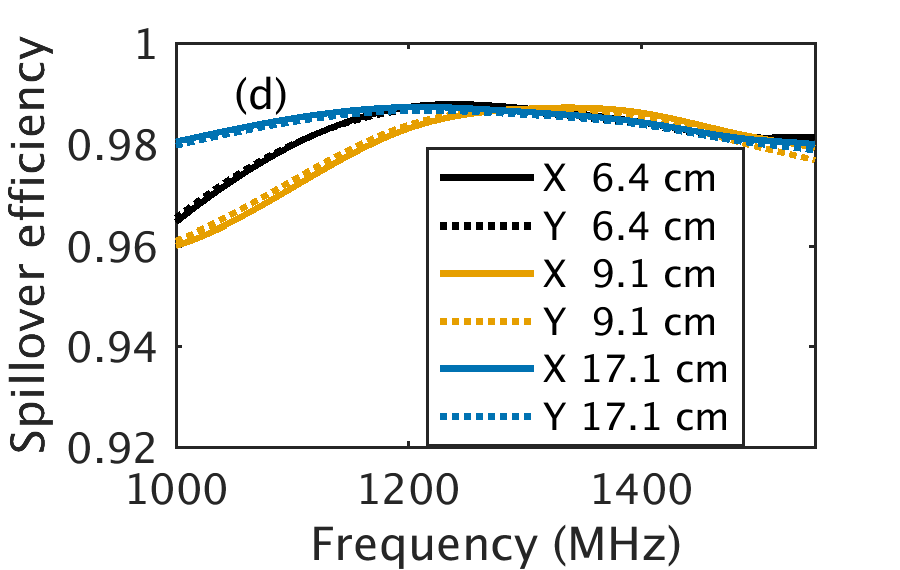}   
\end{tabular}
\caption{(a) The modeled boresight $T_{sys}/\eta$ vs frequency for the
X polarization compared with the measured values taken from Roshi et al. (2018)\cite{roshietal2018}. 
The measured values are same as those shown in Fig.~\ref{fig5}.
The PAF model results for the thermal transition lengths 
$L_{trans} = $ 6.4, 9.1 and 17.1 cm are plotted. 
All model results are obtained with an additional noise contribution of 1 K to 
account for the losses ahead of the LNA.
(b) Same as (a) but for Y polarization. (c) Model X and Y polarization
aperture efficiencies vs frequency
for the three thermal transition lengths. (d) Model X and Y polarization
spillover efficiencies vs 
frequency for the three thermal transition lengths.
}
\label{fig6}
\end{figure}

\begin{figure}[!t]
\centering
\begin{tabular}{c}
\includegraphics[width=3.0in]{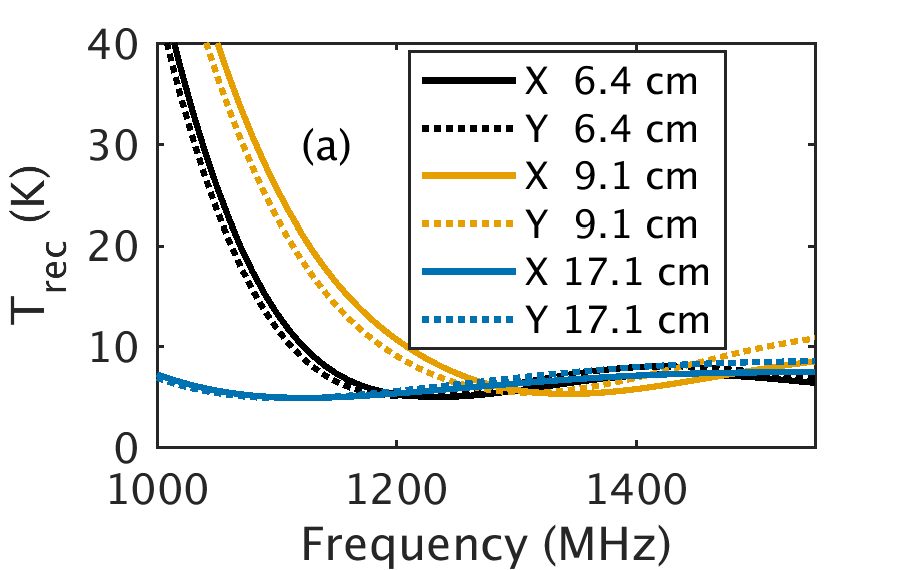} \\
\includegraphics[width=3.0in]{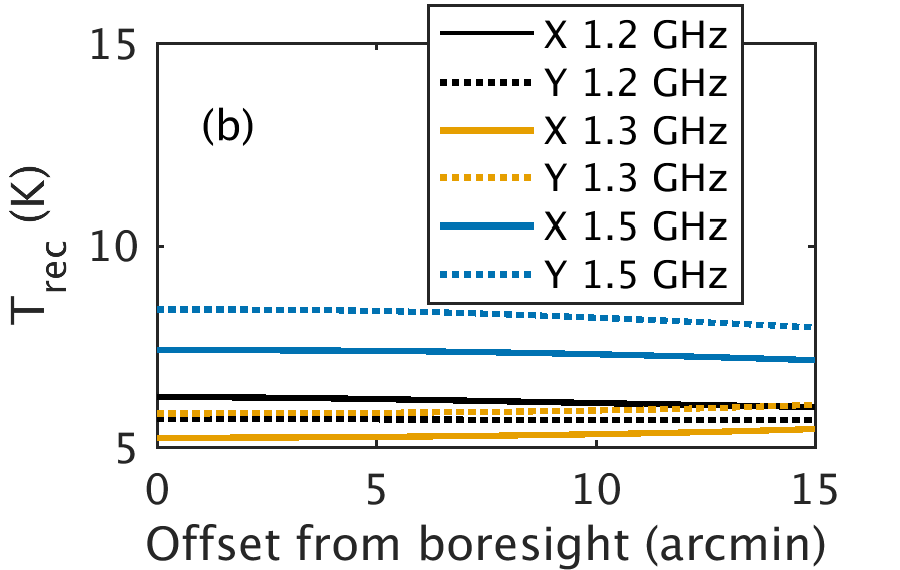}  
\end{tabular}
\caption{(a) Model X and Y polarization receiver temperatures vs frequency 
for three different $L_{trans}$. The receiver temperatures are obtained with Eq.~\ref{youreq}
using the weights that maximizes the SNR.  (b) Model X and Y polarization
receiver temperatures vs offset from
boresight for frequencies 1.2, 1.3 and 1.5 GHz. }
\label{fig7}
\end{figure}

The PAF model $T_{sys}/\eta$ vs frequency for the boresight
direction are obtained for different lengths, $L_{trans}$, for the thermal transition,
which is used as the noise matching network in the model. These results are 
obtained with an additional contribution to noise matrix corresponding to 
$T_{loss}$ = 0, 1, and 2 K to account for the
losses ahead of the LNA (see Section~\ref{loss}). The model results for
$L_{trans}$ = 9.1 cm and for the three values of $T_{loss}$ are shown in Fig.~\ref{fig5}.  
The measured X polarizations $T_{sys}/\eta$ vs frequency for the boresight  
taken from Roshi et al. (2018) \cite[their Fig. 14]{roshietal2018} are included 
in Fig.~\ref{fig5}
for comparison. As seen from Fig.~\ref{fig5}, a reasonable fit to the data points is 
obtained with $T_{loss}$ = 1 K and $L_{trans}$ = 9.1 cm. The physical length
of the thermal transition in the PAF is about 6.4 cm. We attribute the discrepancy 
between the two physical lengths to unaccounted electrical length in the signal path.  

The model predictions for different $L_{trans}$ values for the boresight $T_{sys}/\eta$ vs frequency 
for the X and Y polarizations are shown in Figs.~\ref{fig6}a and \ref{fig6}b, respectively.
The model results are plotted for 3 representative values for 
$L_{trans} = $ 6.4, 9.1 \& 17.1 cm and $T_{loss}$ = 1 K. 
The measured values are also shown in these figures for comparison. 
As seen from Fig.~\ref{fig6}a and \ref{fig6}b, the bandwidth performance of 
the existing system can be increased to $\ge$ 300 MHz by noise matching
the PAF to the LNA. 


The model aperture efficiency, spillover efficiency and receiver temperature
as a function of frequencies are shown in Fig.~\ref{fig6}c, \ref{fig6}d and \ref{fig7}a.
The aperture efficiency obtained is $\sim$ 60\% for the three models near 1350 MHz, 
while the spillover efficiency is $\sim$ 98.7\%. These values are consistent 
with the preliminary model results presented in Roshi et al. (2018) \cite{roshietal2018}.
The beamformed receiver temperature is $\sim$ 5.5 K for the new amplifier model
and for $L_{trans} = 9.1$ cm at 1350 MHz. This receiver temperature is about 2 K smaller
than the earlier result (see \cite{roshietal2018}). 

The reported aperture efficiency of room temperature PAFs on parabolic reflectors
are $>$ 70\% \cite{cappelanbakker2010, chippendaleetal2016}. 
The weights for beamforming were obtained by maximizing the SNR 
for both room temperature PAF observations and our observations.
In such maximization,
the achieved aperture and spillover efficiencies are function of receiver
temperature { (see also \cite{jeffsetal2008})}. 
Generally the aperture efficiency (non-linearly) increases with
increase in receiver temperature. Thus the lower aperture efficiency of $\sim$ 60\%
obtained for our system is a result of lower receiver temperature for the cryogenic
PAF. The aperture
efficiency of the cryogenic PAF is comparable to a corrugated horn with an edge taper
of $\sim$ 22 dB placed at the prime focus of the GBT (S. Srikanth, NRAO, private
communication). The spillover efficiency of the PAF, on the other hand, is a factor
of $\sim$ 1.8 times better than that achieved with the same corrugated horn.  

\subsection{$\frac{T_{sys}}{\eta}$ over the FoV}

\begin{figure}[!t]
\centering
\begin{tabular}{c}
\includegraphics[width=3.0in]{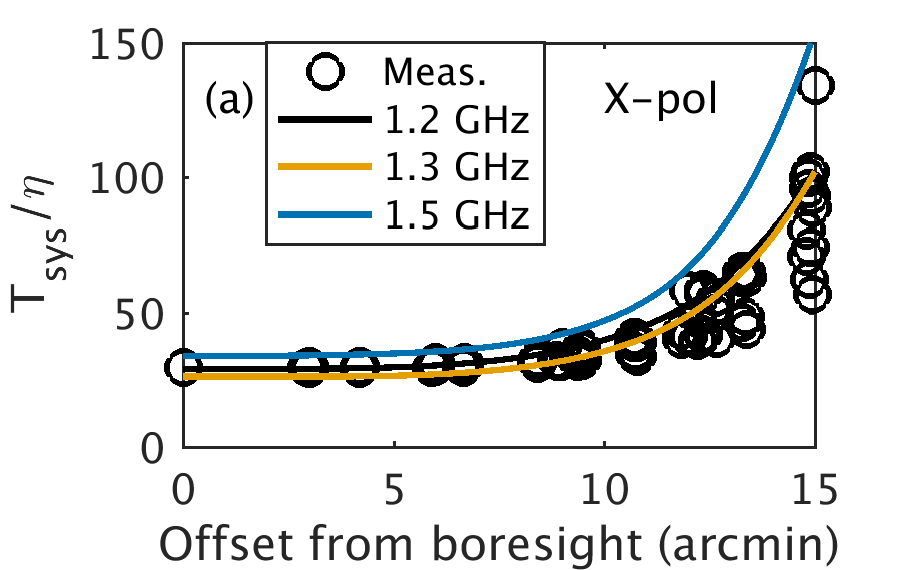} \\
\includegraphics[width=3.0in]{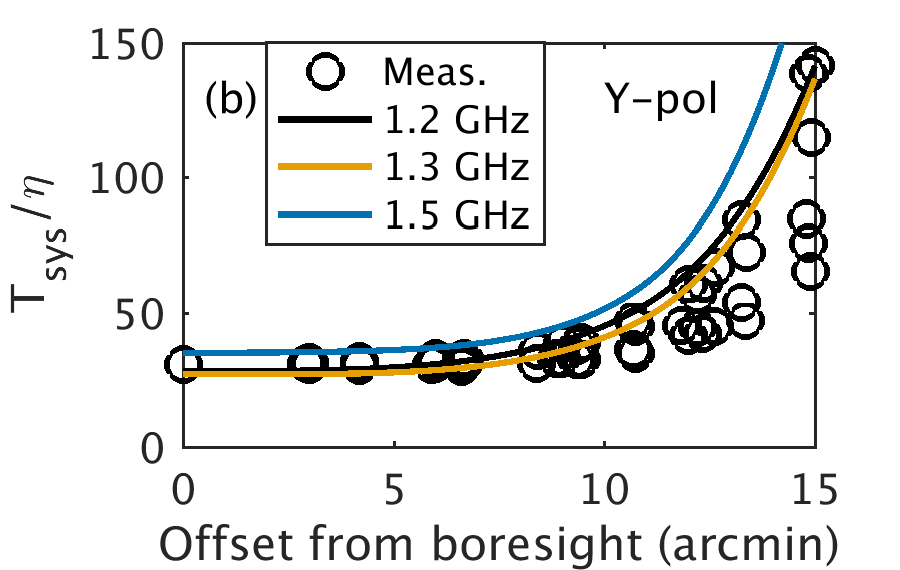} \\
\includegraphics[width=3.0in]{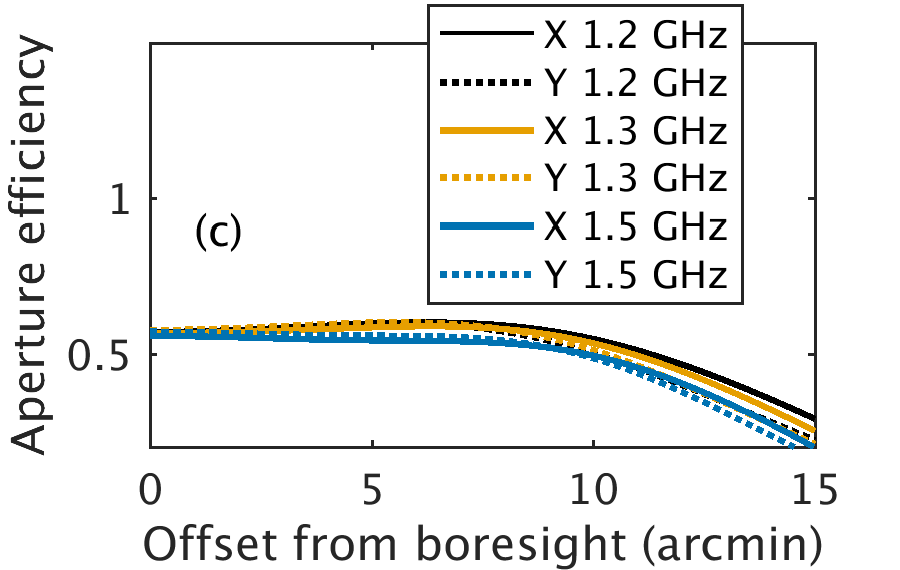} \\
\includegraphics[width=3.0in]{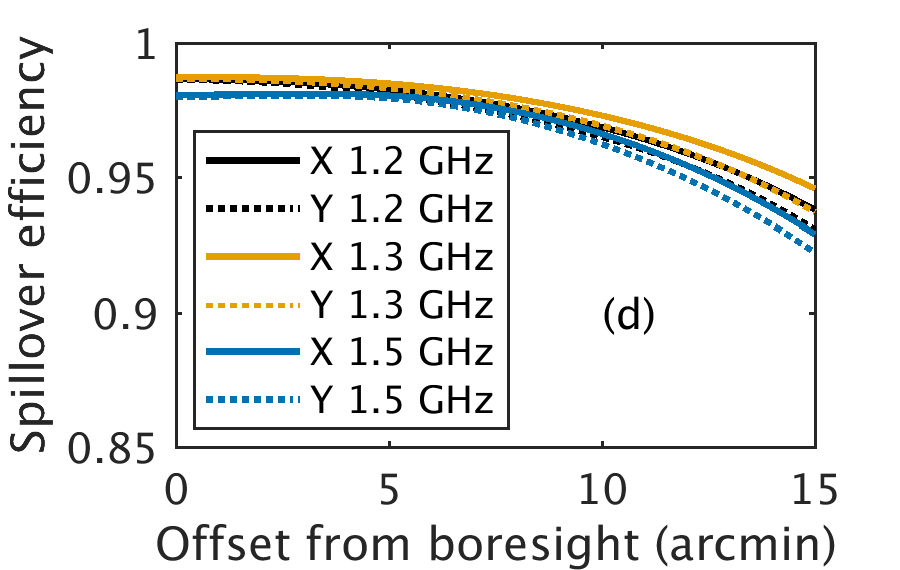}
\end{tabular}
\caption{
(a)The measured and modeled $T_{sys}/\eta$ vs offset from boresight for the
X polarization. The measured values at 1336 MHz (taken from \cite{roshietal2018}) are shown
in circles. The PAF model results for $L_{trans} = 9.1$ cm and $T_{loss}$ = 1K
are plotted for frequencies 1.2, 1.3 and 1.5 GHz.
(b) Same as (a) but for Y polarization. (c) Model X and Y polarization
aperture efficiencies vs offset from boresight 
for the three frequencies. (d) Model X and Y polarization
spillover efficiencies vs offset from boresight 
for the three frequencies.
}
\label{fig8}
\end{figure}

The radial distributions (i.e. offset from the boresight) of the measured 
$T_{sys}/\eta$ for X and Y polarizations along with model results are 
shown in Fig.~\ref{fig8}a and \ref{fig8}b respectively. The measurements are made
at 1336 MHz. The model results for $L_{trans} = 9.1$ cm and $T_{loss}$ = 1 K
are shown for frequencies 1.2, 1.3 and 1.5 GHz. As seen from Fig.~\ref{fig8}a and \ref{fig8}b,
the model results for 1.3 GHz compare well with measurements. 
The model aperture efficiency, spillover
efficiency and receiver temperature as a function of offset from boresight
are shown in Fig.~\ref{fig8}c, \ref{fig8}d and \ref{fig7}b respectively. 
The performance of the system degrades beyond $\sim$ 5\arcmin\ 
because the Airy pattern shifts to the edge of the array and
hence there are not enough elements to form a high sensitivity beam
(see \cite{roshietal2018} for further details). Thus the field
of view of the PAF is limited by the size of the dipole array. 

\subsection{Receiver temperature}

\begin{figure}[!t]
\centering
\includegraphics[width=3.0in]{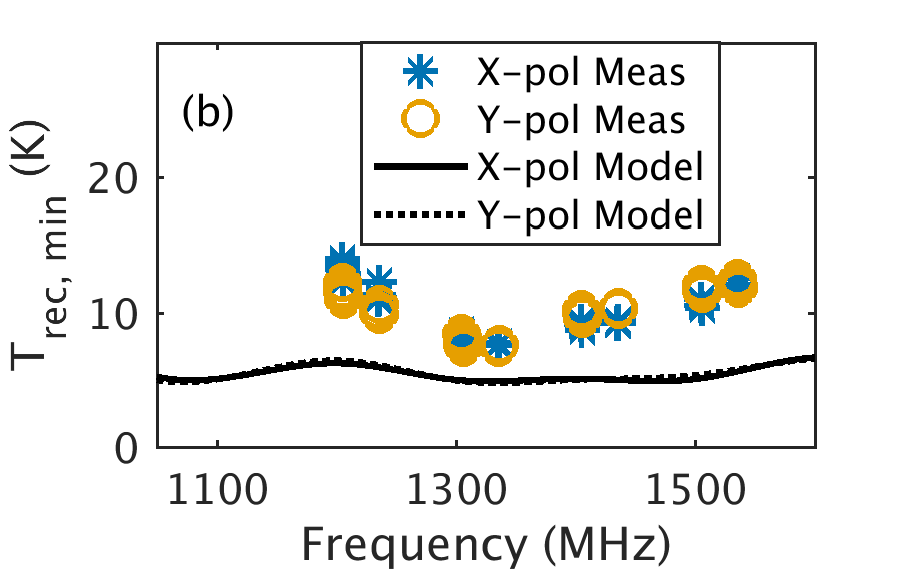}
\caption{
The minimum receiver temperature $T_{rec,min}$ obtained from measurement (
see \cite{roshietal2018})  
and the PAF model vs frequency.
The model results are for $L_{trans} = 9.1$ cm and $T_{loss}$ = 1 K.
The expected uncertainty in the measured $T_{rec,min}$ is $\sim$ 1.1 K near 1350 MHz. }
\label{fig9}
\end{figure}

We compare the model receiver temperature with measurements made
at the outdoor test facility at the GBO. As mentioned in Section~\ref{loss},
these measurements are affected by ground scattered radiation. The
value of the receiver temperature depends on the weights 
used to obtain them (see Eq.~\ref{youreq} and also \cite{roshietal2018}).
The receiver temperature that contributes to the system temperature 
when the PAF is on the telescope can be obtained from the knowledge of
the beamformer weights and complex system gain. However, determining
the on-telescope receiver temperature is currently not 
possible since the measurement
system has different system gain compared to that on the telescope.
Therefore, here we compare model results with 
the minimum receiver temperature, $T_{rec,min}$ (see Section~\ref{loss}).
The reference plane and the length of the thermal transition used for
these comparisons are, respectively, at the input of the thermal transition
(see Fig.~\ref{fig1}b) and $L_{trans} = 9.1$ cm.  
  


We plot $T_{rec,min}$ obtained from
the measurements (see \cite{roshietal2018}) and that obtained using Eq.~\ref{youreq} 
vs frequency in Fig.~\ref{fig9}. 
The model values and measurements differ by $\sim$ 1.5 K near 1350 MHz.
The difference between model values and measurements increases
to $\sim$ 5 K near 1200 and 1500 MHz (see Fig.~\ref{fig9}).  
{ We believe
part of this discrepancy may be due to any uncanceled ground scattering
in the measured $T_{rec,min}$.}
Understanding the discrepancy and investigating methods to improve the receiver temperature 
measurements are ongoing research activity. 

\subsection{Uncertainty in the model results}

The model uses a set of input parameters and results obtained
from electromagnetic simulation of the dipole array. Currently, the uncertainties
in both these quantities are not well determined. Further, the accuracy of
the measured results against which model predictions are compared 
are currently limited by systematic errors.

We expect that the accuracy of the model results is dominated by the uncertainty in the input
parameters to the computational model. The embedded beam patterns and impedance matrix are
obtained from the output of the CST MWS simulation with solver mesh size $\lambda/20$. We
found that our CST simulation results have comparable values with results obtained
from HFSS\footnote{http://www.ansys.com/products/electronics/ansys-hfss} (Warnick, private communication) 
as well as those obtained from the CST with a coarser mesh ($\lambda/10$)
size. However, a direct comparison of the field patterns and S-parameters with
measurements is currently not available. The secondary fields due to reflections
from the telescope surface are computed using a geometric optics approximation,
but we have not quantified the inaccuracies in this computation.  
The uncertainties
in the other input parameters are discussed in Section~\ref{pafsystem}, \ref{loss} \& \ref{modoutput}. 
To obtain an uncertainty for the model results, we estimate the results
by changing the input parameters by an amount equal to their 
uncertainties. The estimated variation in the model values for
$T_{sys}/\eta$ for $L_{trans} = 9.1$ cm
and $T_{loss}$ = 1 K is $\sim$ 15\% near 1350 MHz. This uncertainty in the model result
is valid over a frequency range of $\sim$ 150 MHz centered at 1350 MHz.

\section{Conclusion}
\label{conclu}

In this paper, we have presented a model for noise matched phased array feed system.
The PAF model predictions
were then compared with measurement results. The measurements were made with a 
cryogenic 19 element dual polarized PAF mounted on the GBT. This PAF was designed 
to operate over a frequency range $\ge$ 150 MHz centered at 1350 MHz. Our main results were:
\begin{enumerate}
\item 
We have derived a new expression for the covariance matrices for signal 
by applying the Lorentz reciprocity theorem. An expression for 
the receiver noise covariances matrix and noise temperature were derived 
in terms of the Lange invariants, which were advantageous for noise matching the PAF for
optimum performance. 

\item
We have shown that the PAF model predictions compare well with the measured 
$T_{sys}/\eta$ made on the GBT, both as a function of frequency and 
as a function of offset from boresight. 

\item
Our modeling have indicated that the bandwidth performance of the PAF could be improved by 
a factor of two (i.e. $\ge$ 300 MHz) compared to the current performance
by noise matching the PAF to the LNA.

\item
The PAF model results differ by $\sim$ 5 K (maximum) from the measured 
minimum receiver temperature. 
Investigation to understand this discrepancy and to improve receiver temperature
measurements are underway.
\end{enumerate}

\appendices

\section{}
\label{rec_temp_ap}

The power spectral density due to receiver noise can be converted to a physical temperature
using the definition of noise figure \cite{kerrranda2010} (see also \cite{warnicketal2010}). 
{\em In the absence of a matching network}, the receiver temperature of the PAF 
can be obtained using Eq.~\ref{tn} as \cite{roshifisher2016}
\begin{multline}
T_{rec}  =  T_0 \; \frac{\parbox{2.2in}{$\bm w_1^H \Big(R_n\bs{\mathcal{I}} +  
 \sqrt{R_n g_n} \;\big(\rho \bm Z + \rho^* \bm Z^H\big)$ \\
\hspace*{1.25in} $ + g_n \bm Z\bm Z^H \Big) \;\bm w_1 $}} 
{\frac{1}{2}\bm w_1^H (\bm Z + \bm Z^H) \; \bm w_1}.
\label{trec1}
\end{multline}
Equation~\ref{trec1} can be rewritten as
\bea
&  & \frac{1}{2}\;\bm w_1^H \big(\bm Z + \bm Z^H\big) \bm w_1 \; \frac{T_{rec}}{T_0} \nonumber \\
&  =  & \bm w_1^H \Big(R_n \bs{\mathcal{I}} + \sqrt{R_n g_n} \; \big(\rho \bm Z + \rho^* \bm Z^H\big) + g_n\bm Z\bm Z^H \Big)\bm w_1 \nonumber \\
& = & \bm w_1^H \Big(2 \; \sqrt{R_n g_n \big(1 - \rho_i^2\big)} \; \mbox{Re}\{\bm Z\} + \nonumber \\
&   & \;\;\;\;\;\;\;\; 2 \; \rho_r \sqrt{R_n g_n} \; \mbox{Re}\{\bm Z\}\Big) \; \bm w_1 + \nonumber \\
&   & \bm w_1^H \Big(\frac{R_n g_n}{g_n} \bs{\mathcal{I}} + g_n \bm Z \bm Z^H  - 2 \; \sqrt{R_n g_n \big(1 - \rho_i^2\big)} \; \mbox{Re}\{\bm Z \} \nonumber \\
&   & \;\;\;\;\;\;\;\; - 2 \; \rho_i \sqrt{R_n g_n} \; \mbox{Im}\{ \bm Z \}\Big) \; \bm w_1 \nonumber \\
& = & 2\Big(\sqrt{R_n g_n \big(1 - \rho_i^2\big)}  + \rho_r \sqrt{R_n g_n}\Big) \frac{1}{2}\bm w_1^H \big(\bm Z + \bm Z^H\big) \bm w_1 + \nonumber \\
&   & \bm w_1^H \Big(\frac{R_n g_n}{g_n} \bs{\mathcal{I}} + g_n \bm Z \bm Z^H - 2\; \sqrt{R_n g_n \big(1 - \rho_i^2\big)} \; \mbox{Re}\{\bm Z\} \nonumber \\
&  &  \;\;\;\;\;\;\;\; - 2\;\rho_i \sqrt{R_n g_n} \; \mbox{Im}\{\bm Z\}\Big)\; \bm w_1 \nonumber 
\eea
\bea
& = & \frac{T_{min}}{T_0} \; \frac{1}{2} \; \bm w_1^H \big(\bm Z + \bm Z^H\big)\; \bm w_1 + \nonumber \\
&   & \bm w_1^H \Big(g_n |Z_{LNAopt}|^2 \bs{\mathcal{I}} + g_n \bm Z \bm Z^H - \nonumber \\
&   &  \;\;\;\;\;\;\;\;\;\;\; 2\; g_n \; \mbox{Re}\{Z_{LNAopt}\} \; \mbox{Re}\{\bm Z\} - \nonumber \\
&   &  \;\;\;\;\;\;\;\;\;\;\;  2\; g_n \; \mbox{Im}\{Z_{LNAopt}\} \; \mbox{Im}\{\bm Z\}\Big) \; \bm w_1 \nonumber \\
& = & \frac{T_{min}}{T_0}\; \frac{1}{2}\; \bm w_1^H \big(\bm Z + \bm Z^H\big) \; \bm w_1 + \nonumber \\
&  &  \;\;\;\;\;\;\;\; g_n \bm w_1^H \big(\bm Z - Z_{LNAopt} \bs{\mathcal{I}}\big) \big(\bm Z - Z_{LNAopt}\bs{\mathcal{I}}\big)^H \; \bm w_1 \nonumber \\
& = & \frac{T_{min}}{T_0} \; \frac{1}{2} \; \bm w_1^H \big(\bm Z + \bm Z^H\big) \; \bm w_1 + \nonumber \\
\label{youreq1}
&   & \; N \; \frac{\bm w_1^H \big(\bm Z - Z_{LNAopt} \bs{\mathcal{I}}\big) \big(\bm Z - Z_{LNAopt}\bs{\mathcal{I}}\big)^H\; \bm w_1} {\mbox{Re}\{Z_{LNAopt}\}} \\
& & {\text{Eq.~\ref{youreq1} can be readily rewritten as Eq.~\ref{youreq}.}} \nonumber
\eea

\section{}
\label{symbol_list}
\begin{tabular}{ll}
$A_{ap}$ & Aperture area of the telescope \\
$\bm G $ & Diagonal matrix of overall system gain \\
$g_n, R_n, \rho$ & Noise parameters of the LNA \\
$\bs{\mathcal{I}}$ & Identity matrix \\
$\bm I_0$ & Vector of port currents \\
$k_B$ & Boltzmann constant \\
$L_{trans}$ & Length of the thermal transition \\
$\bm M$ & Characteristic matrix of the PAF system \\
$N$ & Lange invariant \\
$\bm R_{loss}$ & Open circuit voltage covariance (OCVC) matrix \\
              & due to the losses ahead of the LNA \\
$\bm R_{rec}$ & OCVC matrix due to the receiver noise \\
$\bm R_{signal}$ & OCVC matrix due to the source \\
$\bm R_{sky}$ & OCVC matrix due to the sky background \\
              &  and atmospheric noise \\
$\bm R_{spill}$ & OCVC matrix due to the spillover noise \\
$S_{source}$ & Flux density of the source \\
$T_0$ & Reference temperature, 290 K \\
$T_g$ & Ground temperature (300 K) \\
$T_{loss}$ & Temperature due to losses ahead of the LNA \\
$T_{min}$ & Minimum LNA noise temperature \\
$T_{rec, min}$ & Minimum receiver temperature\\
$T_{spill}$ & Spillover temperature \\
$T_{sys}$ & System temperature \\
$\bs{\tilde{V}}$ & Voltage vector at the beamformer input \\
$\bm V_{oc}$ & Open circuit voltage vector \\
$\bm w$ & Weight vector applied at the beamformer \\
$\bm w_1$ & Modified weight vector when using \\
         &  OCVC matrix  \\
$\bm Z$ & Impedance matrix of the PAF \\
$Z_0$ & Free space impedance \\
$z_0$ & Reference impedance, 50 $\Omega$ \\
$Z_{in}$ & Input impedance of the receiver \\
         & system at the reference plane \\
$\bm Z_{in}$ & Diagonal matrix of elements $Z_{in}$ \\
$Z_{LNA}$ & Input impedance of the LNA \\
$Z_{LNAopt}$ & Optimum impedance of the LNA \\
$Z_{opt}$ & Value of $Z_{LNAopt}$ at the reference plane \\
$\eta_{ap}$ & Aperture efficiency \\
$\eta_{spill}$ & Spillover efficiency \\
$\eta$ & Product of $\eta_{ap}$ and PAF's radiation efficiency \\
$\bs{\vec{\mathcal{\psi}}^e}, \bs{\vec{\Psi}^e}$ & Vectors of embedded beam patterns \\
$\bs{\vec{\mathcal{\psi}}^e_{pap}}$ & Vectors of embedded beam patterns \\
          & propagated to the projected aperture plane \\
\end{tabular}
 
\section*{Acknowledgment}

The authors would like to acknowledge very useful discussions and suggestions from Matt Morgan
during the initial phase of the development of the PAF model. The possibility
of an analytical solution for PAF problem was pointed out to D. A. Roshi
by Stuart Hay, CSIRO. The authors acknowledge the efforts of Bob Simon, Steve White and other 
Green Bank Observatory and NRAO Technology Center staff in successfully 
building and testing the PAF receiver. 
We thank Marian Pospieszalski, Anthony Kerr
for useful discussions on the noise properties of the receiver, Srikanth
for discussions on the computation of the GBT aperture field, Wavley Groves and Matt Morgan
for providing the amplifier noise model,  
Robert Dickman and S. K. Pan for the support
and useful discussions during the course of this work.
We thank the referees for providing very useful comments and suggestions,
which has significantly improved the paper.

\ifCLASSOPTIONcaptionsoff
  \newpage
\fi



%

\bibliographystyle{IEEEtran}
\bibliography{mybib}




%

\begin{IEEEbiography}{D. Anish Roshi}
graduated with an engineering degree
in electronics and communications from the University of Kerala in 1988. He 
then joined the Tata Institute of 
Fundamental Research (TIFR), National Center for Radio Astrophysics, Pune as a Research Associate/Scientific Officer.
At TIFR, he completed his Ph.D (Physics) specializing in Radio Astronomy (1999) and was appointed as a member
of the faculty. He was a Jansky postdoctoral fellow at the National Radio Astronomy Observatory (NRAO),
USA between 2000 and 2002. On completion of the post doctoral fellowship, Dr Roshi joined the
Raman Research Institute (RRI), Bangalore.  While at  RRI,  he was involved with several
Astrophysics research and instrumentation projects. He led a team of engineers to build a digital
receiver for the Murchison Widefield Array, located at Western Australia. In 2010, Dr. Roshi moved
to the National Radio Astronomy Observatory (NRAO), USA as a member of the scientific staff. At NRAO,
he was the System Architect and Project Scientist for the Versatile Astronomical Spectrometer that
was built for the Green Bank Telescope (GBT). He was also the System Architect of the Phased Array
Feed/Beamformer project. He is currently a Senior Observatory Scientist for Radio Astronomy
at the Arecibo Observatory. His research interests include Galactic interstellar medium, 
massive star formation and radio astronomy instrumentation.
\end{IEEEbiography}


\begin{IEEEbiography}{William Shillue} is a
Research Engineer at the National Radio Astronomy Observatory (NRAO), Central Development Laboratory (CDL), Charlottesville, VA. He received the BS degree in electrical engineering from Cornell University in 1985, and an MS degree in electrical engineering from Univ. of Massachusetts in 1990. In 1991 he joined the National Radio Astronomy Observatory (NRAO) in Green Bank, WV, and participated in the design and deployment of a 13.7-m satellite earth station in support of the NASA Orbiting-Very-Long-Baseline-Interferometer Project. From 1994-2005 he worked at NRAO Tucson developing millimeter-wave receivers and instrumentation for the 12-m telescope, and new photonics technologies for the Atacama Large Millimeter Array (ALMA). From 2005-2012, he was team leader for the ALMA Central Local Oscillator and ALMA photonic LO distribution systems. More recently, he has led the NRAO Phased Array Research and Development program and developed technology concepts for next generation radio astronomy instrumentation. He is a member of IEEE Photonics Society and Microwave Theory and Techniques Society. 
\end{IEEEbiography}


\begin{IEEEbiography}{J. Richard Fisher}
was born in Pittsburgh, PA in 1943 but he
spent his school years on a small farm near Reynoldsville in
western Pennsylvania.  He received a BS in physics from Penn State
in 1965 and a PhD in astronomy from the University of Maryland in
1972.  He joined the NRAO scientific staff at Green Bank, WV
immediately following graduate school and has held various
positions such as Head of the Electronics Division, Site Director,
and Project Manager of several instrumentation projects.  In 1978
he took leave from the NRAO to spend 18 months at the Division of
Radio Physics, CSIRO in Australia and 3 months at the Raman
Research Institute in Bangalore, India.  He moved to the Charlottesvile,
VA offices of NRAO in 2005 and held the position of Chief Technologist
until his retirement in 2012.  He is now Scientist Emeritus at NRAO.
His research interests include cosmology, antenna design, and signal processing.

\end{IEEEbiography}







\end{document}